\documentclass[preprint]{aastex}

\shorttitle{Kerr Accretion Flows}
\shortauthors{De Villiers, Hawley, and Krolik}

\begin{document}

\title{Magnetically Driven Accretion Flows in the Kerr Metric
I: Models and Overall Structure}


\author{Jean-Pierre De Villiers, John F. Hawley}
\affil{Astronomy Department\\
University of Virginia\\ 
P.O. Box 3818, University Station\\
Charlottesville, VA 22903-0818}
\and
\author{Julian H. Krolik}
\affil{Physics and Astronomy Department\\
Johns Hopkins University\\ 
Baltimore, MD 21218}
\email{jd5v@virginia.edu; jh8h@virginia.edu; jhk@pha.jhu.edu}

\begin{abstract}

This is the first in a series of papers that investigate the properties
of accretion flows in the Kerr metric through three-dimensional,
general relativistic magnetohydrodynamic simulations of tori with a
near-Keplerian initial angular velocity profile.  We study four models
with increasing black hole spin, from $a/M=0$ to $0.998$, for which the
structural parameters of the initial tori are maintained nearly
constant.  The subsequent accretion flows arise self-consistently from
stresses and turbulence created by the magnetorotational instability.
We investigate the overall evolution and the late-time global structure
in the resulting non-radiative accretion flows, including the magnetic
fields within the disks, the properties of the flow in the plunging
region, and the flux of conserved quantities into the black hole.
Independent of black hole spin, the global structure is described in
terms of five regions: the main disk body, the coronal envelope, the
inner disk (consisting of an inner torus and plunging region), an
evacuated axial funnel, and a bi-conical outflow confined to the
corona-funnel boundary.  We find evidence for lower accretion rates,
stronger funnel-wall outflows, and increased stress in the near hole
region with increasing black hole spin.

\end{abstract}


\keywords{Black holes - magnetohydrodynamics - instabilities - stars:accretion}

\section{Introduction}

With the recognition that the magneto-rotational instability (MRI;
Balbus \& Hawley 1991, 1998)  drives disk turbulence and the outward
transport of angular momentum, the theory of black hole accretion is
moving from relatively simple parametrized, one-dimensional, and
time-independent models toward genuine dynamics.  Because of the
complexity of the underlying physics, however, large-scale computer
simulations are essential.  With the ever-increasing power of
computers, such simulations are now possible, and their realism should
rapidly improve.  The goal is increasingly detailed and predictive
theoretical models to compare with and provide interpretation for
observations.

Until recently, most simulations of accretion flows were carried out
using Newtonian gravity, or a pseudo-Newtonian approximation to the
Schwarzschild potential.  In particular, a number of pseudo-Newtonian
simulations have focused on the inner edge of an accretion disk
orbiting a black hole (Hawley \& Krolik 2001, 2002, hereafter HK01,
HK02; Armitage, Reynolds \& Chiang 2001; Armitage \& Reynolds 2003;
Machida \& Matsumoto 2003).  The dynamical behavior of the inner edge
is critical in determining the efficiency of accretion as well as the
energy and angular momentum carried into the hole.  Recent analytic
work (Krolik 1999; Gammie 1999) and numerical studies have found that, in
contrast to the conventional picture, magnetic stress can operate
on the flow within the last stable orbit, increasing the nominal
efficiency of the accretion and delivering less specific angular
momentum to the hole.  Simulations
have also shown that the turbulence in the disk is marked by rapid
and large amplitude fluctuations, resulting in a highly nonsteady
accretion rate.  The implication of these results for black hole
accretion disks were summarized by
Krolik \& Hawley (2002; hereafter KH02) who identified several distinct
functional definitions of the inner edge of an accretion disk.

While the general properties of an accretion disk around a
Schwarzschild black hole may be modeled reasonably well by a
pseudo-Newtonian potential, many of the important features of the inner
edge of disks depend strongly on the spin of the black hole,
characterized by the parameter $a/M$.  Since it is hoped that
observable accretion diagnostics may provide a means to measure $a/M$
in black hole systems, it is essential to understand the specific
effects of black hole spin, and this requires simulations using full
relativity.  In this study, we model accretion flows using ideal
general relativistic magnetohydrodynamics (GRMHD) in the static
background of a Kerr black hole with the recently developed code of
De Villiers \& Hawley (2003; hereafter DH03a).

In an earlier paper, De Villiers \& Hawley (2003; hereafter DH03b)
presented an initial survey of the properties of accretion tori in the
Kerr metric at moderate resolution.  These first simulations used
constant specific angular momentum (constant-$l$) tori threaded by
poloidal loops of magnetic field as the initial state.  These
simulations demonstrated similarities between accretion into a
Schwarzschild hole and earlier pseudo-Newtonian simulations of Hawley
(2000; hereafter H00), while explicitly highlighting the differences
that result from the Kerr metric.  The effect of frame dragging was
observed, especially in the retrograde torus where it caused a reversal
of flow direction in the inspiraling accretion stream.  The black hole
spin determines the location of the marginally stable orbit, which, in
turn, strongly affects the innermost regions of the flow.  A constant
angular momentum torus is a particularly short-lived configuration in
the presence of the MRI, and its initial evolution includes a violent
transient accretion phase as the disk readjusts to a more nearly
Keplerian angular momentum distribution.  After this, the disk models
settled into a state of sustained turbulence with an accretion rate
that decreased with increasing $a/M$.  In all models, the magnetic
pressure became strong in the near-hole region, with a late-time
magnetic pressure comparable to the initial gas pressure maximum in the
torus.

In this paper, we pick up the thread of HK01 and HK02 and present the
results of a series of high-resolution simulations of nonradiative
accretion flows in the Kerr metric.  We revisit the initial state of
model GT4 originally investigated in H00, and subsequently the focus of
HK01 and HK02, which features an initial angular momentum distribution
with a radial dependence that is slightly shallower than
Keplerian.   The torus is seeded with loops of weak poloidal magnetic
field.  As in the earlier non-relativistic studies, we will pay
particular attention to the inner region of the accretion disk since
the most extreme relativistic effects and the greatest energy release
happen near the radius of the marginally stable orbit, $r_{ms}$.  In
the near-hole region, an accretion flow must transition from strong
turbulence within the disk to a more nearly laminar flow as matter
plunges toward the black hole, and simulations find that the location
of this transition is highly time-variable.  Generally the transition
begins outside of $r_{ms}$, but the transonic plunging inflow is not
fully established until some point inside of $r_{ms}$.  The inner disk
edge is not abrupt:  average flow variables, including the stress, are
relatively smooth, albeit time-varying, functions of radius.  Magnetic
stress continues to be important even within the plunging region.  In
fact, the magnetic stress becomes proportionally more important as
flux-freezing increases the field strength relative to the gas pressure
(Krolik 1999), and increases the correlation between the radial and
azimuthal field components.

This paper is the first in a series that will investigate in detail
a coherent set of accretion disk simulations.  Here
we introduce the basic models and present a preliminary analysis of the
results.  In \S2 we describe the analytic expressions that are used to
initialize the models, and the parameters of the four models studied.
We also list a number of the physics diagnostics used in the
simulations.  In \S3 we present an overview of results of the
three-dimensional global MHD simulations, focusing on the broad
evolutionary features, and aspects that are common to all four models.
We examine some of the late-time properties of the disks, highlighting
the key features through a structural classification of the accretion
system. In this classification, we distinguish five main regions:  the
main disk body, the coronal envelope,  the inner disk, the funnel-wall
jet, and the axial funnel.  In \S4 we review and summarize our
findings.  In the course of the presentation and analysis of results,
we pose some of the questions that will be the subject of subsequent
investigations.

\section{Numerical Approach and Initial Conditions}

We use a finite difference code to solve the equations of ideal GRMHD
in the spacetime of a Kerr (rotating) black hole.  We adopt the
familiar Boyer-Lindquist coordinates, $(t,r,\theta,\phi)$, for which
the line element has the form, 
\begin{equation}\label{kerr}
{ds}^2=g_{t t}\,{dt}^2+2\,g_{t \phi}\,{dt}\,{d \phi}+g_{r r}\,{dr}^2+
 g_{\theta \theta}\,{d \theta}^2+g_{\phi \phi}\,{d \phi}^2 .
\end{equation}
We use the metric signature $(-,+,+,+)$, along with geometrodynamic
units where $G = c = 1$.  Time and distance are in units of the black 
hole mass, $M$.  The determinant of the 4-metric is $g$, and 
$\sqrt{-g} = \alpha\,\sqrt{\gamma}$, where $\alpha$ is the lapse function,
$\alpha=1/\sqrt{-g^{tt}}$, and $\gamma$ is the determinant of the
spatial $3$-metric. We follow the usual convention of using Greek characters
to denote full space-time indices and Roman characters for
purely spatial indices.

The equations of ideal GRMHD are the
law of baryon conservation, ${\nabla}_{\mu}\,(\rho\,U^{\mu}) = 0$,
where ${\nabla}_{\mu}$ is the covariant derivative, the conservation of
stress-energy, ${\nabla}_{\mu}{T}^{\mu\,\nu} = 0 $, where
${T}^{\mu\,\nu}$ is the energy-momentum tensor for the fluid, and the
induction equation, ${\nabla}_{\mu}{}^*{F}^{\mu\,\nu}= 0$, where
${}^*F^{\mu \nu}$ is the dual of the electromagnetic field strength
tensor.  
The precise form of the equations that are solved in the code
is derived in DH03a, along with a set of primary and secondary
variables especially suited to numerical evolution.  A succinct
presentation of the equations is provided in DH03b, and will not be
repeated here, although for reference we reiterate the definitions of
the primary code variables.  

The state of the relativistic test fluid at each point in the spacetime
is described by its density, $\rho$, specific internal energy,
$\epsilon$, $4$-velocity $U^\mu$, and isotropic pressure, $P$.  The
relativistic enthalpy is $h=1 + \epsilon + P/\rho$.  The pressure is
related to $\rho$ and $\epsilon$ through the equation of state of
an ideal gas, $P=\rho\,\epsilon\,(\Gamma-1)$, where $\Gamma$ is the
adiabatic exponent.  For these simulations we take $\Gamma=5/3$.  The
entropy of the gas can increase in regions of strong compression
(shocks), through the action of an artificial viscosity.

The magnetic field of the fluid is described by two sets of variables,
the constrained transport magnetic field, $F_{jk}={\cal{B}}^i$, and
magnetic field $4$-vector $\sqrt{4\pi} b^\mu = {}^{*}F^{\mu\nu}U_\nu$.  
The ideal MHD condition requires
$U^\nu F_{\mu\nu} = 0$.  The magnetic field $b^\mu$ is included in
the definition of the total four momentum,
\begin{equation}\label{momdef}
 S_\mu = (\rho\,h\ + {\|b\|}^2)\,W\,U_\mu ,
\end{equation}
where $W$ is the boost factor.  We
define auxiliary density and energy functions $D = \rho\,W$ and $E =
D\,\epsilon$, and transport velocity $V^i = U^i/U^t$.  
We also define the specific angular momentum as
$l=-U_\phi/U_t$ and the angular velocity as $\Omega = U^\phi/U^t$.
The numerical scheme 
is built on the set of
variables $D$, $E$, $S_\mu$, $V^i$, ${\cal{B}}^i$, and $b^\mu$.

\subsection{Torus Initial State}

For consistency with past studies we choose as our initial condition a
somewhat thick torus with a nearly-Keplerian distribution of angular
momentum,  specifically a general relativistic version of the GT4 model
of H00.  To create such a model we follow the procedure described by
Chakrabarti (1985) to construct equilibria with non-constant angular
momentum distributions in the Kerr metric.  We seek a solution where
the disk has a power-law rotation, 
\begin{equation}\label{kd.1} 
\Omega = \eta\,\lambda^{-q} 
\end{equation} 
where $\eta$ is a constant, $q$ is
a positive number, and $\lambda$ is given by
\begin{equation}\label{kd.7}
\lambda^2 = {l \over \Omega} 
 = l {\left(g^{t\,t}-l\,g^{t\,\phi} \right) \over 
      \left(g^{t\,\phi}-l\,g^{\phi\,\phi} \right)} .
\end{equation}
In the Schwarzschild metric, this parameter has the form
$\lambda^2 = -g^{t\,t}/g^{\phi\,\phi}$.  In Newtonian gravity
$\lambda$ is simply the cylindrical radius.

In the hydrodynamic limit, the momentum evolution equation is 
(Hawley, Smarr, \& Wilson 1984)
\begin{equation}\label{kd.2}
\partial_t\left(S_j\right)+
  {1 \over \sqrt{\gamma}}\,
  \partial_i\,\sqrt{\gamma}\,\left(S_j\,V^i\right)+
  {1 \over 2}\,\left({S_\epsilon\,S_\mu \over S^t}\right)\,
  \partial_j\,g^{\mu\,\epsilon}+
  \alpha\,\partial_j\left(P\right) = 0 ,
\end{equation}
where $S_j = \rho\,h\,W\,U_j$ is the momentum, and the other variables are 
defined as above.
The momentum equation is simplified by imposing time-independence, 
axisymmetry, and requiring that there be no poloidal motion.  This 
simplified equation reads,
\begin{equation}\label{kd.3}
  \alpha\,\partial_j\left(P\right)+
  {1 \over 2}\,\left({S_\epsilon\,S_\mu \over S^t}\right)\,
  \partial_j\,g^{\mu\,\epsilon} = 0 .
\end{equation}
Using the definition of the momentum $4$-vector, and the definition
of specific angular momentum, we obtain
\begin{equation}\label{kd.4}
  {\partial_j\left(P\right) \over \rho\,h} = 
 -{{U_t}^2 \over 2}\,\partial_j\left({U_t}^{-2}\right)+
  {U_t}^2\,\left(-\partial_j\,g^{t\,\phi}+l\,\partial_j\,g^{t\,\phi} \right)
  \,\partial_j\,l,
\end{equation}
where ${U_t}^{-2}=g^{t\,t}-2\,l\,g^{t\,\phi}+l^2\,g^{\phi\,\phi}$,
a result which follows from $U_\mu U^\mu = -1$ and $U_r = U_\theta = 0$.

Under the assumption of constant entropy, $T ds = 0$, we use  $dh=dp/\rho$
to write
\begin{equation}\label{kd.5}
  {\partial_j\,h \over h}= 
 -{1 \over 2}\,{\partial_j\left({U_t}^{-2}\right)\over {U_t}^{-2}}+
  {U_t}^2\,\left(-\partial_j\,g^{t\,\phi}+l\,\partial_j\,g^{t\,\phi} \right)
  \,\partial_j\,l.
\end{equation}
In order to integrate this equation, we assume that $\Omega \equiv \Omega(l)$,
and use the definition of ${U_t}$ above, along with the relation
$\Omega = \left(g^{t\,\phi}-l\,g^{\phi\,\phi} \right)/
\left(g^{t\,t}-l\,g^{t\,\phi} \right)$ to get
\begin{equation}\label{kd.6}
  {\partial_j\,h \over h}= 
 -{1 \over 2}\,{\partial_j\left({U_t}^{-2}\right)\over {U_t}^{-2}}+
  {\Omega \over 1- l\,\Omega}\,\partial_j\,l .
\end{equation}
(As pointed out by Abramowicz {\it{et al.}} (1978), $-U_t(r,\theta;l)$
plays a role analogous to an effective potential.)
In order to find a disk solution, we
use (\ref{kd.1}) and the definition of $\lambda$ to write 
$l = \Omega \lambda^2 = \eta\,\lambda^{2-q}$ and 
$\Omega = \eta^{-2/(q-2)}\,l^{q/(q-2)} \equiv k\,l^\alpha$ . So,
\begin{equation}\label{kd.6a}
  {\int}_{h_{in}}^{h} {dh \over h}= 
 -{1 \over 2}\,{\int}_{U_{in}}^{U_t} {d\left({U_t}^{-2}\right)\over {U_t}^{-2}}+
  {\int}_{l_{in}}^{l} {k\,l^\alpha \over 1- k\,l^{\alpha+1}}\,d\,l
\end{equation}
where $\alpha = q/(q-2)$ and 
$x_{in}$ refers to the quantity in question evaluated on the surface
of the disk. Clearly, $h_{in}=0$, and a general solution admitting a
choice of surface binding energy $U_{in}$ is given by
\begin{equation}\label{kd.10}
  h(r,\theta) = { U_{in} f(l_{in}) \over U_{t}(r,\theta) f(l(r,\theta))} ,
\end{equation}
where $f(l) = {\|1 - k\,l^{\alpha+1}\|}^{1/(\alpha+1)}$ or
$f(\Omega) ={\|1 - k^{-1/\alpha}\,\Omega^{(\alpha+1)/\alpha}\|}^{1/(\alpha+1)}$.
Using the equation of state and the definition of enthalpy, the
internal energy of the disk is
\begin{equation}\label{kd.11}
  \epsilon(r,\theta) = {1 \over \Gamma} \left(
  { U_{in} f(l_{in}) \over U_{t}(r,\theta) f(l(r,\theta))}-1\right) .
\end{equation}
For a constant entropy adiabatic gas the pressure is given by $P =
\rho\,\epsilon\,(\Gamma - 1) = K\,\rho^\Gamma$, and density is given by
$\rho={\left[{\epsilon\,(\Gamma - 1) / K}\right]}^{1/(\Gamma - 1)}$.

For the Schwarzschild metric, these analytic relations completely
specify the initial equilibrium torus.  In the Kerr metric, $\lambda^2$
is an implicit function of $l$, requiring an iterative procedure to
solve for the equilibrium state.  However, it is not essential to
construct a torus that is in strict hydrostatic equilibrium.  It is
sufficient to use the Schwarzschild expression for $\lambda$ to
initialize an approximate equilibrium for a torus orbiting a Kerr
hole.  The differences between even an extreme Kerr and a Schwarzschild
metric are relatively small at the location of the initial pressure
maximum for the tori of interest here.  In addition, we introduce a
seed magnetic field and ignore its contribution to the equilibrium
solution, which is also a satisfactory approximation as long as the
imposed field is weak.  In the final analysis, motions resulting from
the MRI will quickly dominate the evolution regardless.

A particular disk solution is specified by choosing the parameter $q$,
the entropy parameter $K$, and the angular momentum $l_{in}$ at
$r_{in}$, the inner edge of the disk.  For all simulations, the
location of the inner edge, as well as parameters $K=0.01$ and $q=1.68$
are kept fixed.  The specific angular momentum $l_{in}$ is chosen to
maintain the location of the pressure maximum constant for all
values of the Kerr parameter $a/M$.

The initial magnetic field is obtained from the definition of $F_{\mu
\nu}$ in terms of the $4$-vector potential, $A_\mu$,
$F_{\mu \nu} = \partial_\mu A_{\nu} - \partial_\nu A_{\mu}$ .
Our initial field consists of axisymmetric poloidal field loops, laid
down along isodensity surfaces within the torus by defining
$A_{\mu} = (A_t,0,0,A_\phi)$, where
\begin{equation}\label{vecpot}
A_\phi = 
\cases{
k (\rho-\rho_{cut}) & for $\rho \ge \rho_{cut}$ \cr
0 & for $\rho < \rho_{cut}$},
\end{equation}
where $\rho_{cut}$ is a cutoff density corresponding to a particular
isodensity surface within the torus.  Using the above definition, it
follows that ${\cal{B}}^r = -\partial_\theta A_{\phi}$ and
${\cal{B}}^\theta = \partial_r A_{\phi}$.  The constant $k$ is set by
the input parameter $\beta$, the ratio of the gas pressure to the
magnetic pressure, using the volume-integrated gas pressure divided by
the volume-integrated magnetic energy density in the initial torus.  We
use $\beta=100$ in all runs except for one model (KDP) which instead uses 
$\beta=200$.
The constant $\rho_{cut}$ is chosen to keep the initial magnetic field
away from the outer edge of the disk.  Here we use $\rho_{cut} = 0.5
\rho_{max}$, where $\rho_{max}$ is the maximum density at the center of
the torus, to ensure that the initial field loops are confined well
inside the torus.

The region outside the torus is initialized to a numerical vacuum that
consists of a cold, tenuous, non-rotating, unmagnetized gas. The
auxiliary density variable, $D$, in the vacuum is set to seven orders
of magnitude below the maximum value of $D$ in the  initial torus.
Similarly, the auxiliary energy variable $E$ is set fourteen orders of
magnitude below the initial maximum of $E$. These values define the
numerical floor of the code, below which $D$ and $E$ are not allowed to
drop. In practice, the numerical floor is rarely asserted during a
simulation, since outflow from the evolving torus quickly populates
the grid with a gas that, though of low density, lies above the
numerical floor. Further details on the numerical floor are given in
DH03a and DH03b.

\subsection{Overview of Torus Models}

In this series of simulations we concentrate on the effect of different
black hole spin parameters upon the evolution of similar initial torus
models.  We consider a torus orbiting a Schwarzschild black hole, and
tori orbiting prograde Kerr holes with values $a/M=0.5$, $0.9$
and $0.998$.  These models are designated KD0 (Kepler disk, 0 black
hole spin), KDI (intermediate spin), KDP (strong prograde spin), and
KDE (extreme spin) respectively.  We choose parameters for the initial
tori that keep the inner edge of the disk and the location of the
initial pressure maximum constant as the black hole rotation parameter
is varied.  Table \ref{params} lists the general properties of the
models, where $l_{in}$ denotes the specific angular momentum at
$r_{in}$, the inner edge of the disk (in the equatorial plane),
$(-U_{t})_{in}$ is the energy per unit mass of matter at the surface of
the torus, ${r_P}_{max}$ is the location of the pressure maximum (also in
the equatorial plane), and $T_{orb}$ is the orbital period at the pressure
maximum in units of $M$.  For reference we also list $r_{ms}$, the
location of the marginally stable orbit, the values of $(U_\phi )_{ms}$
and $(U_t)_{ms}$ for a particle in a circular orbit at $r_{ms}$
(Bardeen, Press, \& Teukolsky 1972), and $T_{orb\,(ms)}$, the orbital
period of a test particle at the marginally stable orbit in units of
$M$.  All radii in the table are in units of $M$. Note that the
quantity $T_{orb}$ evaluated at the pressure maximum is used in the
following sections as the unit of the evolution time of the
simulations.

\begin{table}[ht]
\caption{\label{params}Global Torus Simulation Parameters.}
\begin{tabular}{lcrrrrrcrrrr}
 & & & & & & & & & &\\
\hline
Model & $a/M$ & $\beta$ & $l_{in}$ & $\left(U_t\right)_{in}$ & $r_{in}$ & ${r_P}_{max}$ & 
$T_{orb}$ & 
$r_{ms}$ & $\left( U_\phi\right)_{ms}$ & $\left( U_t\right)_{ms}$ & 
$T_{orb\,(ms)}$\\
\hline
\hline
KD0 & 0.000 & 100 & 4.66 & -0.9725 & 15.0 & 25.0 & 785 & 6.00 & 3.464 & -0.943
& 97.95 \\
KDI & 0.500 & 100 & 4.61 & -0.9730 & 15.0 & 25.0 & 794 & 4.23 & 2.903 & -0.918
& 62.57 \\
KDP & 0.900 & 200 & 4.57 & -0.9732 & 15.0 & 25.0 & 803 & 2.32 & 2.100 & -0.844
& 31.04 \\
KDE & 0.998 & 100 & 4.57 & -0.9734 & 15.0 & 25.0 & 803 & 1.23 & 1.392 & -0.679
& 15.73 \\
\hline
\end{tabular}
\end{table}

Both moderate and high-resolution simulations were carried out for
these models.  The lower-resolution simulations use a $128 \times 128
\times 32$ grid in $(r,\theta, \phi)$; these models are indicated with
an ``lr'' appended to their names, e.g., KD0lr.  The high-resolution
simulations use a $192 \times 192 \times 64$ grid.  The azimuthal grid
spans the quarter plane, $0 \le \phi \le \pi/2$, with periodic boundary
conditions in $\phi$.  The $\theta$-grid ranges over $0.045\, \pi \le
\theta \le 0.955\, \pi$, with an exponential grid spacing function that
concentrates zones near the equator.  Reflecting boundary conditions
are enforced in the $\theta$-direction.  The outer radial boundary is
set to $r_{max}=120 M$ in all cases; the inner radial boundary is
located just outside the horizon, with the specific value depending on
the location of the horizon as determined by the Kerr spin parameter
$a/M$. The inner boundary is at $r_{min}= 2.05\,M$, $1.90\,M$,
$1.45\,M$, and $1.175\,M$ for models KD0, KDI, KDP, and KDE,
respectively. The radial grid is set using a hyperbolic cosine function
to maximize the resolution near the inner boundary. The hyperbolic
cosine function ensures that there is a tighter grid spacing ($\Delta
r_i$) near the inner boundary than would be provided by the exponential
spacing in traditional ``logarithmic'' grids. This type of grid ensures
that the plunging region is well resolved in the KD0, KDI, and KDP models, and marginally so in model KDE where this
region occupies a very small range of radii near the inner boundary.
Boundary values are obtained for all variables, including magnetic
fields, by extrapolation into the boundary with a slope of zero.  To
prevent flow from the boundaries onto the grid, 
the radial momentum, $S_r$, is set to zero in the ghost
zones if the flow direction would otherwise be onto the grid.  The
boundary values for the 
transport velocities, $V^i$, are constructed from the boundary momenta
using the usual momentum normalization procedure (DH03a).

\begin{figure}[ht] 
\epsscale{1.0}
   \plotone{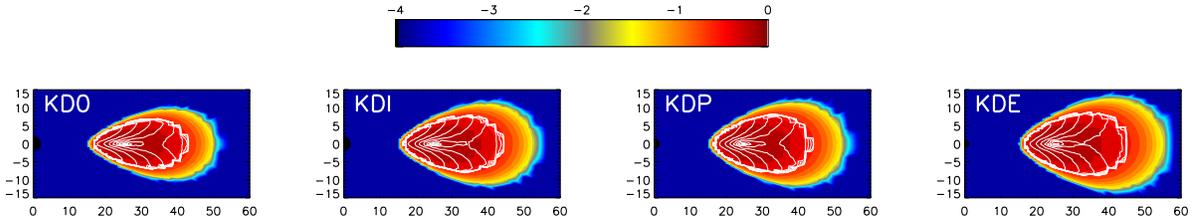}
   \caption{\label{KD_init} 
    Initial density ($\rho$) and magnetic pressure (${\|b\|}^2/2$)
    profiles for the KD0, KDI, KDP, and KDE models. Density is
    scaled logarithmically in relation to the maximum, as shown
    in the color bar.
    The magnetic pressure, shown as $10$ overlaid contour lines, is 
    also scaled logarithmically over $4$ decades from the 
    pressure maximum. The individual plots are
     labeled by model.  }  
\end{figure}

Figure \ref{KD_init} shows the initial density profiles for these tori,
plotted on the same spatial scale to illustrate clearly the relative
size of the disks.  The color map saturates in the vacuum region
surrounding the disk, where the density ($\rho$) is seven
orders of magnitude below $\rho_{max}$.  Note that choosing to fix the
inner edge and pressure maximum locations in the tori means that the
greater the spin parameter, the larger the torus.  Overlaid on the tori
are contours of the initial magnetic pressure plotted on a logarithmic
scale.  The choice of vector potential density cut-off (eq.
\ref{vecpot}), $\rho_{cut} = 0.5\,\rho_{max}$, yields initial poloidal
loops concentrated around the pressure maximum, and fully contained
within the torus.  The chevron-shaped magnetic pressure contour lines
show that the magnetic pressure is largest inside ${r_P}_{max}$.  The
field in this region is rapidly amplified by shear, and this
contributes to the expulsion of a thin stream of magnetized gas through
the inner edge of the disk during the first orbit.

\subsection{Evolution Diagnostics}\label{diags}

Full three-dimensional time-dependent simulations can generate an
enormous amount of data.  To handle this data, we have developed a
variety of evolution diagnostics designed to organize and analyze the
results.  First, complete dumps of the code variables are saved at
periodic intervals, and these can be examined in detail after a
simulation.  For these models such dumps are saved every $80\,M$ in
time; this represents $10$ dumps per orbital period at ${r_P}_{max}$
or, equivalently, $0.2$ to $1.2$ dumps per orbital period at $r_{ms}$,
depending on $a/M$.  In addition, we save the density variable every
$2\,M$ in time in order to produce high time-resolution animations.  We
also compute a more extensive set of azimuthally-averaged variables,
variables averaged on spherical shells, and volume-integrated
quantities.  These are saved every $1\,M$ of time.  
All this data is archived and will serve as a
resource for more detailed studies in subsequent work.

At each radius we define the shell-averaged quantity, ${\cal X}$, as
\begin{equation}\label{avgdef}
\langle{\cal X}\rangle(r) = {1 \over {{\cal A}}(r)} \int\int{ 
 {\cal X}\,\sqrt{-g}\, d \theta\,d \phi}
\end{equation}
where the area of a shell is ${\cal{A}}(r)$ and the bounds of integration 
range over the full $\theta$ and $\phi$ computational domains.  
For these simulations we compute shell-averaged values of density,
$\langle\rho\rangle$, density-weighted angular momentum,
$\langle \rho\, l\rangle$, gas pressure, $\langle P\rangle$,  
magnetic field strength $\langle{\|b\|}^2\rangle$, canonical
angular momentum density $\langle \rho\, h \, U_\phi\rangle$
and energy density $\langle -\rho\, h \, U_t\rangle$.  
From these we derive quantities such as the density-weighted average 
specific angular momentum,
$\langle l\rangle =\langle \rho\, l \rangle/\langle\rho\rangle$. 

Fluxes through the shell are computed in a similar manner, but are
not normalized with the area.  We evaluate the rest mass flux
$\langle\rho\,U^r\rangle$, energy flux 
\begin{equation}
\langle{T^r}_t\rangle = \langle{\rho\,h\,U^r\,U_{t}}\rangle
+\langle{{\|b\|}^2\,U^r\,U_{t}}\rangle-\langle{b^r\,b_t }\rangle,
\end{equation}
and the angular momentum flux
\begin{equation}
\langle{T^r}_{\phi}\rangle = \langle{\rho\,h\,U^r\,U_{\phi}}\rangle+
\langle{{\|b\|}^2\,U^r\,U_{\phi}}\rangle-\langle{b^r\,b_\phi}\rangle.
\end{equation}
To provide flexibility in analysis, each of the three components in the 
above sums is calculated and stored separately.  Again, various quantities 
can be subsequently derived from these fluxes and shell averages.

Volume-integrated quantities are computed using
\begin{equation}\label{3avgdef}
\left[{\cal Q}\right] = \int\int\int{ 
 {\cal Q}\,\sqrt{-g}\, dr\,d \theta\,d \phi}.
\end{equation}
The volume-integrated quantities computed and saved as a function of
time are the total rest mass, $\left[ \rho
U^t \right]$, angular momentum $\left[ {T^t}_{\phi}\right]$, and total
energy $\left[ {T^t}_{t}\right]$. 

These diagnostic calculations can be used to monitor the conservation
of rest mass, energy, and momentum during the numerical evolution.  Since
the numerical grid is finite in extent, the conservation laws, when
written in integral form (Page \& Thorne 1974), yield volume and
surface terms that are be evaluated as described above.
Within the limits of sampling frequency, it is possible to
account for change in the total mass, energy, and angular momentum
in terms of the net fluxes that leave the grid.
Specifically, for the integrated quantities $\left[{\cal Q}\right]$, 
we monitor the corresponding cumulative flux through the inner and outer 
radial boundaries,
\begin{equation}\label{surfflux}
\left\{{\cal F}\right\}_{{\rm{in},\rm{out}}} = \int{ 
 dt\,\langle{\cal F}\rangle}(r_{\rm{in},\rm{out}}).
\end{equation}
The history sum for rest mass conservation is
\begin{equation}\label{Mflux}
M_{\rm{tot}} = \left[\rho\,U^t\right]+
 \left\{\rho\,U^r\right\}_{\rm{in}}+
 \left\{\rho\,U^r\right\}_{\rm{out}};
\end{equation}
for energy conservation, it is
\begin{equation}\label{Eflux}
E_{\rm{tot}} = \left[{T^t}_t\right]+
 \left\{{T^r}_t\right\}_{\rm{in}}+
 \left\{{T^r}_t\right\}_{\rm{out}};
\end{equation}
and for momentum conservation ($\phi$-component), it is
\begin{equation}\label{Jflux}
L_{\rm{tot}} = \left[{T^t}_\phi\right]+
 \left\{{T^r}_\phi\right\}_{\rm{in}}+
 \left\{{T^r}_\phi\right\}_{\rm{out}}.
\end{equation}

\section{Synopsis of Results}

\begin{figure}[ht]
    \epsscale{1.0}
    \plotone{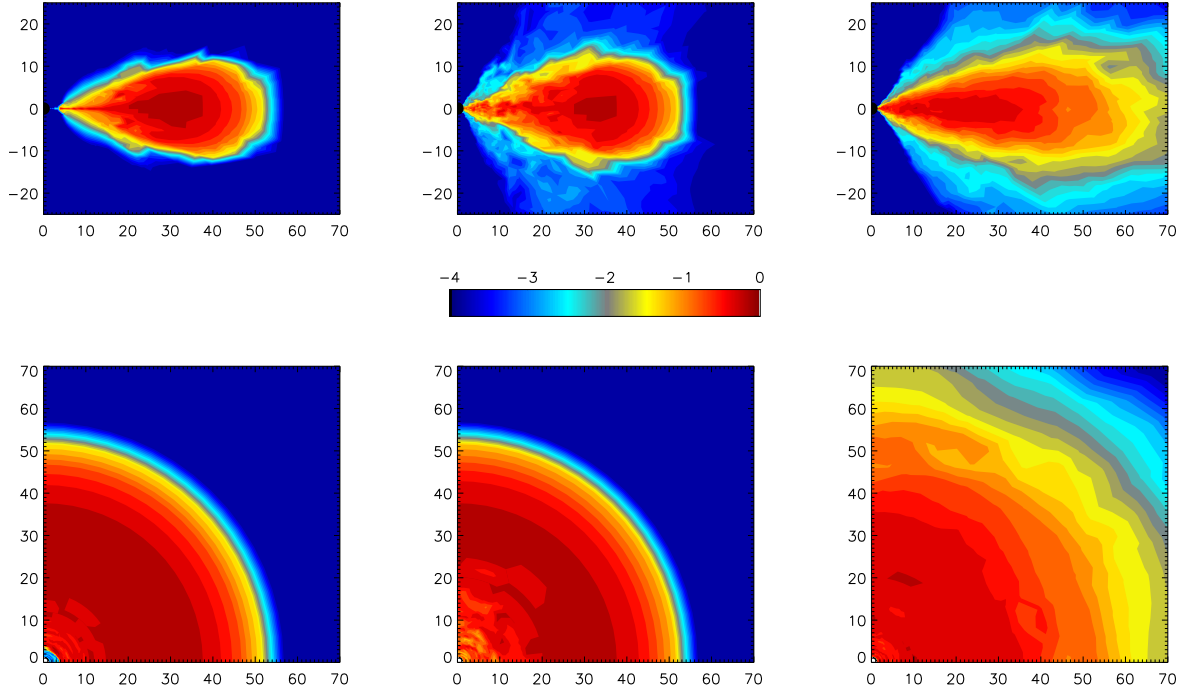}
    \caption{\label{KDPhr} 
     Plots of log density ($\rho$) for model KDP. The top panels
     are polar slices through the disk at $\phi=0$. The bottom panels 
     are equatorial slices through the disk at $\theta=\pi/2$. 
     Left panels are taken at $t=1.0$ orbits;
     center panels at $t=2.0$ orbits;
     right panels at $t=10.0$ orbits.} 
\end{figure}

The four models, KD0, KDI, KDP, and KDE, were evolved for a time
equivalent to $10$ orbits at their respective pressure maxima.  Figure
\ref{KDPhr} shows polar and equatorial slices of the density ($\rho$)
for model KDP at $t=1.0$, $2.0$, and $10.0$ orbits; the figures for the
other three models are comparable, and are not shown.  Overall, the
general evolutionary phases identified in the constant-angular momentum
(SF) models of DH03b apply equally well to the evolution of the KD
models:  (1) the initial rapid growth of the toroidal magnetic field by
shear; (2) the initial nonlinear saturation of the poloidal field MRI,
which progresses outward through the torus; and (3) quasi-stationary
evolution by sustained MHD turbulence in a disk with a roughly constant
opening angle.  However, there are also some differences between the
two types of model. While the SF models experienced a violent transient
prior to the onset of turbulence, there is no comparable event in the
KD models. Instead, there is a gradual migration of a spiraling inflow
towards the black hole (left panels), and the onset of the MRI occurs
fairly quickly in the near-hole region, once the flow is established
(center panels).  This is because the growth rate of the MRI is
proportional to $\Omega$, which increases inward. The onset of the MRI
then progresses gradually outward until the entire disk becomes
turbulent.  Another difference is that the SF models featured a violent
expulsion of large, buoyant magnetic bubbles, and such events are
largely absent in the KD models.   Instead, smaller-scale, low-density,
high-magnetic pressure regions can be seen moving outward along the
surface of the disk (center top panel).  It is likely that the vigor of
the early stages of the evolution in the SF tori derives from the much
stronger inward-directed pressure gradient in the inner half of the
torus.  This pressure remains even after the MRI has cut the
centrifugal support by reducing the super-Keplerian $l$ distribution to
near-Keplerian.  The large, buoyant magnetic bubbles in the SF models
are the result of initial conditions; there is a relatively large
amount of non-magnetized gas surrounding the initial magnetized region
inside the torus and a substantial interchange instability feeds off
this contrast as the field strengthens.

In the late stages of the evolution, all  of the KD models achieve a
quasi-steady state with sustained turbulence.  
The general features observed in the four disk models are summarized in
Figure \ref{simsum}, which shows the time- and azimuthal-average of the
density over the tenth orbit of the KDP simulation.   We distinguish
five separate regions in the late-time flow:  the main body of the
disk, the coronal envelope, the inner disk, consisting of the inner torus 
and plunging region, the funnel-wall jet, and the axial funnel region.  The 
overall characteristics of these regions are as follows:

\begin{figure}[ht]
    \epsscale{0.7}
    \plotone{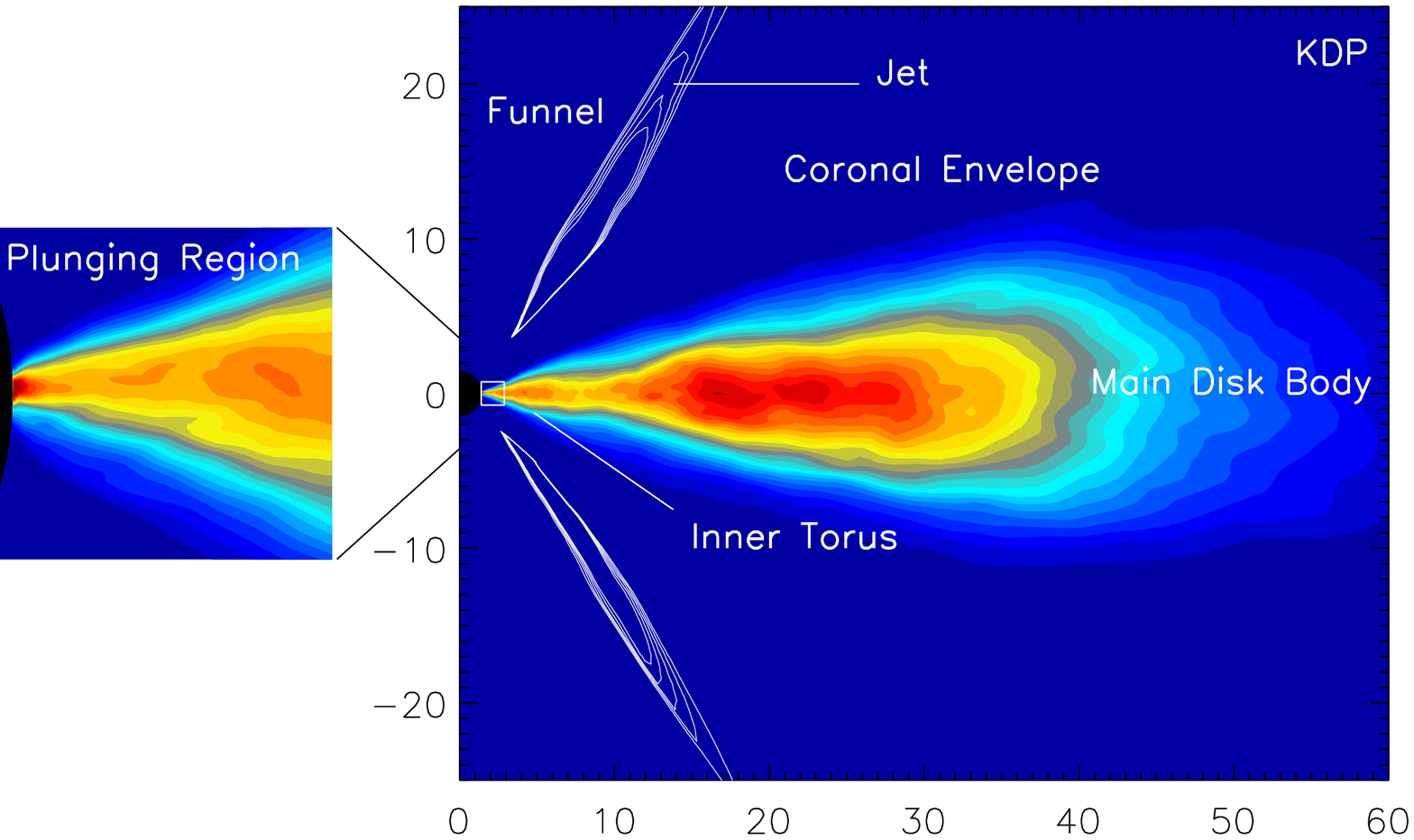}
    \caption{\label{simsum} 
     The azimuthally-averaged density ($\rho$) time-averaged over
     the tenth orbit is plotted on a 
     linear scale for model KDP.  The main dynamical
     features of the system are labeled.  The jet is outlined by contours of
     positive radial momentum.  The box on the left is a close up
     showing the plunging region from just outside the marginally stable
     orbit to the black hole horizon.  The labeled features are seen in
     all of the simulations.  } 
\end{figure}

\begin{itemize}

\item[(1)]  The {\bf main body of the disk} is a turbulent wedge of roughly
constant opening angle containing most of the mass. 
The pressure scale height $H$ at each radius can be measured by
computing the point where the total pressure has fallen to $e^{-1}$ of
its value at the equator.   Measured at $r=20M$, $H/R \sim 0.18$ in
model KD0, $H/R \sim 0.2$ for models KDI and KDP, and $H/R \sim 0.25$
for model KDE.  Inside of $r=20M$, $H/R$ increases systematically with
increasing $a/M$ up $H/R \sim 0.4$ for KDE.  Inside the disk body gas
pressure dominates ($\beta > 1$) and the magnetic and velocity fields
are highly tangled.  The outer part of the disk moves radially outward
with time as it gains angular momentum.

\item[(2)] The {\bf coronal envelope} is a region of low density above
the surface of the disk, where gas and magnetic pressure are comparable
($\beta \sim 1$).  The magnetic field in this region is more regular
than in the disk body.   Animation sequences show an outflow
driven from the inner region of the disk out along the surface of the
disk.  This outflow supplies gas and magnetic field to the coronal
envelope.

\item[(3)] The {\bf inner torus and plunging region} are located at the
inner edge of the accretion disk.  The inner torus is a time-variable
structure where matter accumulates from accretion from the
main disk.  It is marked by a local pressure and density maximum
surrounded by a thickened disk.  This region was referred to as the
``mini-torus'' in DH03b, and the ``inner-torus'' in the
pseudo-Newtonian accretion simulation of Hawley \& Balbus (2002;
hereafter HB02); we adopt the latter term here.  The density and
pressure maximum is located just outside the marginally stable orbit;
hence the inner torus is found at smaller radius with increasing
$a/M$.  In the same way, the inner torus becomes denser (in relation to
the initial density maximum) and hotter with increasing black hole
spin.  The density and pressure drop inside of the inner torus as the
disk approaches the plunging region where matter leaves the disk and
spirals into the black hole.  The plunging region begins near the
marginally stable orbit.  Here matter spirals in toward the event
horizon, stretching field lines as it falls.  Most of the matter in the
inner torus will eventually accrete into the black hole, but some is
ejected into outflows in the coronal envelope and along the funnel
wall.

\item[(4)] The {\bf funnel-wall jet} is an outflow along the centrifugal
barrier which originates in the vicinity of the inner torus.  The density 
in the jet is small compared to the disk, but one to 
two orders of magnitude greater than in the funnel itself.
The contours in Figure \ref{simsum} that delineate the jet are
selected positive values of the time- and azimuthal-average of radial
momentum.  The intensity of the jet is model-dependent:  the jet is
weak in model KD0; there is progressively more outflow with increasing
$a/M$.

\item[(5)] The {\bf axial funnel} is a magnetically-dominated region in
which there is very tenuous gas, several orders of magnitude below
the density in the disk body.  The small amount of gas in the funnel
has negligible angular momentum.  The funnel contains a predominantly
radial magnetic field; the magnetic pressure near the hole is
comparable to the maximum gas pressure in the initial torus.  The
funnel region is cleanly separated from the coronal envelope by the
centrifugal barrier.

\end{itemize}

The entire flow is characterized by large fluctuations in both space
and time.  In addition, animations of the simulation data show
large-scale trailing spiral waves in all regions of the disk and
coronal envelope.

In the following sections we use this structural classification 
of the accretion system as a guide and expand on some of the properties, 
both structural and dynamical, of these regions.

\subsection{The Main Disk Body}\label{disk}

\begin{figure}[ht]
    \epsscale{1.0}
    \plotone{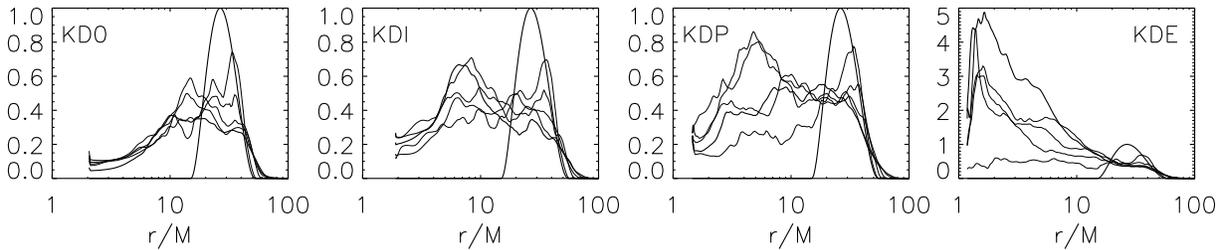}
    \caption{\label{rhoseq} 
     Sequence of density profiles $\langle \rho \rangle(r)$
     for the high-resolution
     models at $t=0$, $2$, $4$, $6$, $8$, and $10$ orbits.  Density
     is given in units of the initial density maximum, $\rho_{max}(t=0)$.
     The thick lines correspond to the initial and final density 
     profiles. The individual plots are labeled by model.  } 
\end{figure}

The disk evolves from its initial state to its late-time quasi-steady
state by redistributing angular momentum and spreading out radially in
both directions.  The evolution is clearly illustrated through a
time-series of shell-averaged densities, $\langle \rho \rangle(r)$, which
are shown in Figure \ref{rhoseq}.  These graphs capture the overall
redistribution of mass, while showing how the inner region of the
accretion disk is established.  The increase in maximum density and the
inward migration of the inner torus with increasing black hole spin are
especially clear in these figures.  

\begin{figure}[ht]
    \epsscale{0.6}
    \plotone{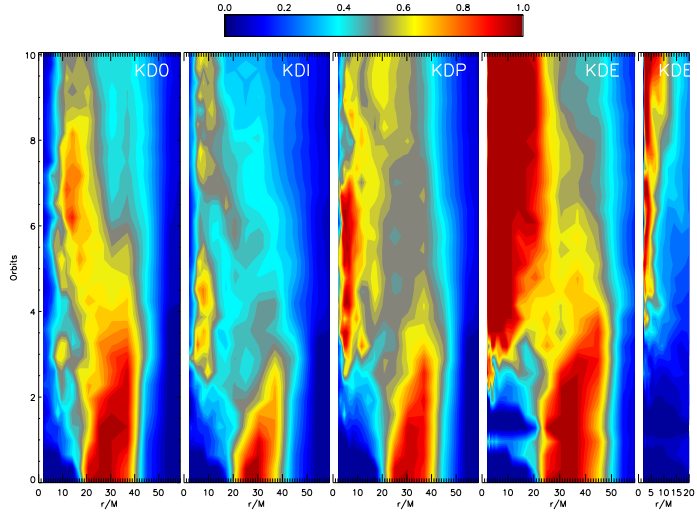}
    \caption{\label{rhost} 
     Spacetime diagrams of $\langle{\rho}\rangle(r,t)$ for high-resolution
     models KD0, KDI, KDP, and KDE.
     Density is plotted on a linear scale, in units of the initial density 
     maximum.
     Dark red and dark blue colors denote saturation of the color
     map. The individual plots are
     labeled by model. The plot on the extreme right shows the 
     inner region 
     of the KDE plot with $r < 20\,M$ where the color scale is 
     saturated in the full
     KDE plot to the left.  Here the linear scale is set to 
     the density maximum of the inner torus.} 
\end{figure}

A more complete summary of the density evolution is provided by the
spacetime diagrams of $\langle{\rho}\rangle (r,t)$ in Figure
\ref{rhost}.  These diagrams illustrate several key points.  First, the
density maximum of the initial torus is rapidly eroded away from the
inside as accretion begins.  This density maximum is effectively gone
by $t \simeq 3-5$ orbits as matter is redistributed within the disk by
the outward transport of angular momentum.  Second,
one can see the development of the inner torus as a density maximum in
the inner disk, fed by accretion from larger radii.  The overall
accretion is marked by intermittent maxima that appear as a series of
leftward-traveling yellow-grey streams.

\begin{figure}[ht]
    \epsscale{1.0}
    \plotone{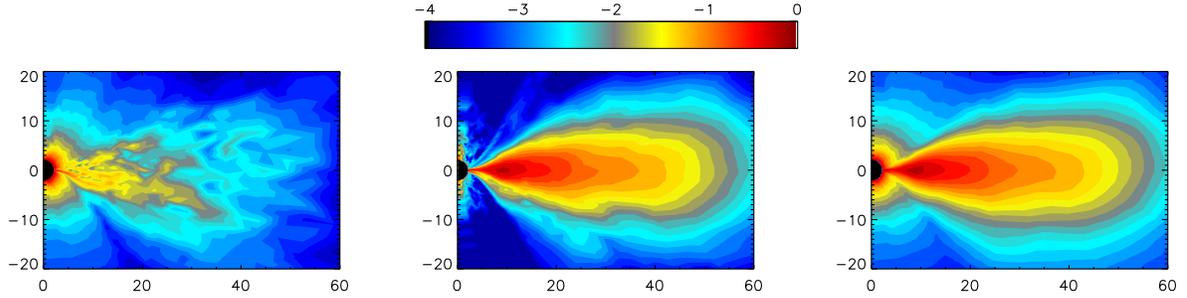}
    \caption{\label{KD0pazavg} 
     Plots of the azimuthally averaged (left) magnetic, (center) gas, and 
     (right) total
     pressure for high-resolution model KD0 at $t = 10$ orbits. 
     Pressure is normalized to maximum total pressure, and
     scaled logarithmically, as shown in the color bar.
     } 
\end{figure}

The distribution of gas pressure $\langle p \rangle (r)$ at late-time
is similar to that of the density for all models.  Of particular
interest is the relative distribution of gas and magnetic pressures.
Figure~\ref{KD0pazavg} shows the late-time, azimuthally averaged value
in model KD0 of gas (left panel), magnetic (center panel), and total
pressure (right panel).  The total pressure distribution is smooth; the
contours are disk-shaped outside the centrifugal barrier, where angular
momentum plays a role in the gas dynamics, and spherically symmetric
inside the funnel where the angular momentum is effectively zero.
Comparison of magnetic and gas pressure with the total pressure shows
that gas pressure dominates inside the disk, magnetic pressure
dominates inside the funnel and in the plunging region, and both 
contribute on comparable terms within the coronal envelope.

\begin{figure}[ht]
    \epsscale{1.0}
    \plotone{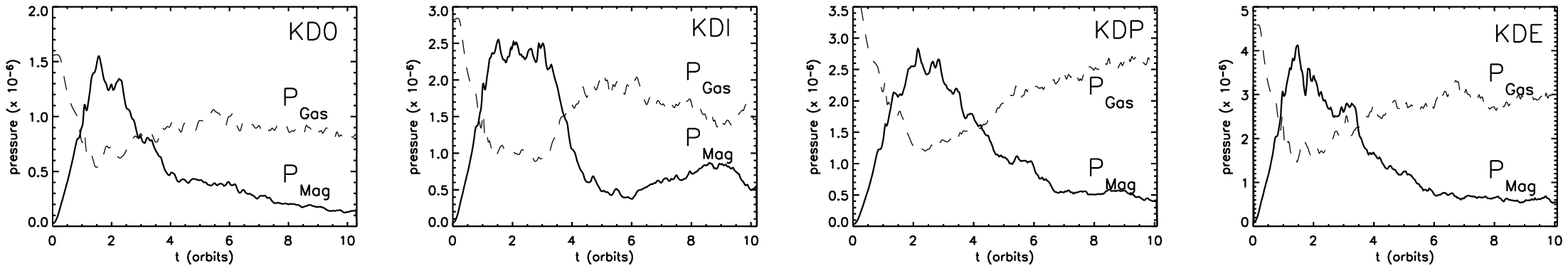}
    \caption{\label{KD_Pmg_rpmax} 
     Gas and magnetic pressure history, $\langle{P_{gas}}\rangle(t)$
     and $\langle{P_{mag}}\rangle$(t), at the location of the initial
     pressure maximum for the high-resolution models.  Pressure is 
     given in code units for all 4 models, allowing cross comparison of
     values.  The individual plots 
     are labeled by model.} 
\end{figure}

Figure \ref{KD_Pmg_rpmax} shows gas and magnetic pressure
time-histories at the radius of the initial gas pressure maximum.  As
with density, gas pressure at this location decreases to a lower
late-time mean value once sustained turbulence is established.
Magnetic pressure, $\langle P_{mag}\rangle$, follows a different
course.  The magnetic field is initially weak and concentrated at the
gas pressure maximum.  As the magnetic field is amplified by shearing
and the MRI, the radius of the magnetic pressure maximum migrates
inward with the accreting matter.  There is a brief period, when the
MRI saturates, when the magnetic pressure exceeds the gas pressure at
the location of the initial gas pressure maximum.  But in the late-time
quasi-steady state of all four models, gas pressure dominates over
magnetic pressure in the main disk.

\begin{figure}[ht]
    \epsscale{1.0}
    \plotone{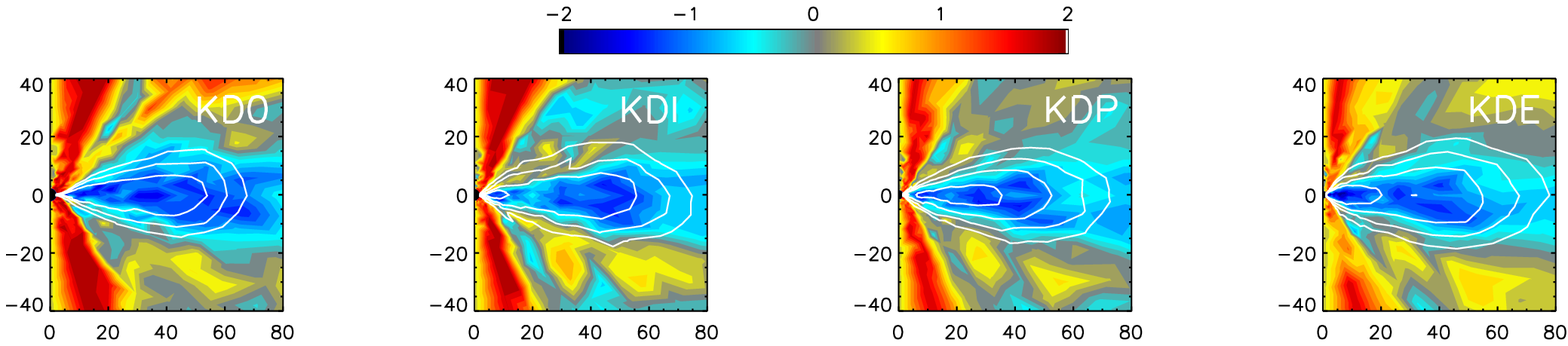}
    \caption{\label{KDpratio} 
     Color contours of
     the ratio of azimuthally averaged magnetic to gas pressure, 
     $P_{mag}/P_{gas}$, 
     for high-resolution models at $t = 10$ orbits.  The scale is
     logarithmic, and is the same for all panels;  
     the color maps saturate in the axial funnel.
     The body of the accretion disk is identified with overlaid
     density contours
     at ${10}^{-2}$, ${10}^{-1.5}$, ${10}^{-1}$, and 
     ${10}^{-0.5}$ of ${\rho}_{max} (t=0)$.
     The individual plots are labeled by model.  } 
\end{figure}

Another view of the pressure distribution within the models is provided
by Figure \ref{KDpratio}, which shows the ratio of magnetic to gas
pressure, $\beta^{-1}$, for each of the high-resolution models. The
color scale is centered on $\beta^{-1}=1$ and shows that the main disk
is gas-pressure dominated, while in the funnel the converse is true.
The coronal envelope stands out in the graph as a region where the gas
and magnetic pressures are comparable but not uniformly distributed;
more will be said about this in \S \ref{corona}.

Full spacetime diagrams of the pressure (not shown) reveal
high-frequency waves throughout the disk, especially in models KDP and
KDE.   The generation of pressure waves seems to be a natural
consequence of the turbulence within the disk.  The most prominent of
these waves seem to occur at frequencies comparable to that of the
accretion streams undergoing the final plunge into the black hole.
They originate near the point where the flow begins to switch from
turbulence-dominated to plunging inflow, a location defined as the
turbulence edge in KH02.  The pressure waves are common to MRI
turbulence, and have been seen in many previous simulations.
Comparable waves were seen in the previous fully relativistic models of
DH03b as well as the pseudo-Newtonian simulations of HK01 and HK02
(see, e.g., Fig. 9 of HK01).  However, with the increased spatial and
temporal resolution of the KD models, a greater wealth of detail is now
available.

\begin{figure}[ht]
    \epsscale{1.0}
    \plotone{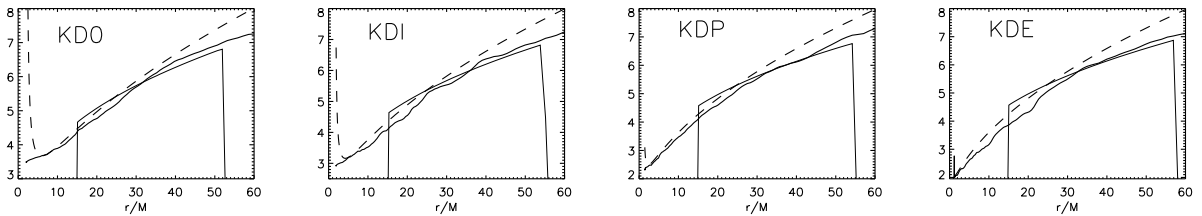}
    \caption{\label{KDel} 
     Specific angular momentum $\langle l \rangle$ as a function of
     radius at $t=0$ (thin line) and
     at $t=10.0$ orbits (thick line).  The individual plots are
     labeled by model.  In each case the Keplerian distribution
     for a test particle, $l_{kep}(r,a)$, is shown as dashed 
     line.} 
\end{figure}

One of the most important dynamical properties of an accretion disk is
the overall distribution of angular momentum.  Figure \ref{KDel} shows
density-weighted specific angular momentum histories $\langle
l\rangle=\langle\rho\, l\rangle/\langle\rho\rangle$ for the four
models.  The KD models begin with a radial distribution of specific
angular momentum that is slightly shallower than Keplerian, shown here
as a thin line.  The late-time angular momentum profiles are shown as
thick lines, and, for reference, the dashed lines depict
$l_{kep}(r,a)$, the Keplerian distribution for test particles.  Since
the models are initialized with a near-Keplerian profile, the overall
change in $l(r)$ is less dramatic than for the SF models of DH03b.
However, the final state in the main disk is similar to the one
found in these models, as well as to the final state found in the  
pseudo-Newtonian simulations of HK01 and HK02, namely a near-Keplerian
slope with a slightly sub-Keplerian value that rises to very close to the
Keplerian value near $r_{ms}$.  Inside the main disk,
contours of constant-$\Omega$ are nearly vertical; in other words, the 
matter is very close to rotating on cylinders. In all models, the profile 
continues to decrease smoothly past $r_{ms}$. 

\subsection{The Coronal Envelope}\label{corona}

The coronal envelope extends outward from the surface of the disk to
the outer boundary of the computational domain and is bounded on the
inside by the centrifugal funnel.  The gas in this region has low
density (Fig. \ref{KDPhr}), but retains substantial specific angular
momentum.  As noted in \S\ref{disk}, the overall magnetic pressure in
the coronal envelope is comparable to gas pressure ($\beta \sim 1$),
although typically the magnetic field is not uniformly distributed in
the coronal envelope.  There are regions that are more or less
magnetized, with $\beta$ ranging from $\approx 0.3$ to $3$.  As a
general rule, the main body of the disk is gas-pressure dominated and
the funnel region is magnetically dominated, so plots of $\beta$
provide a useful means of visualizing the coronal envelope
(Fig. \ref{KDpratio}).

\begin{figure}[ht]
    \epsscale{1.0}
    \plotone{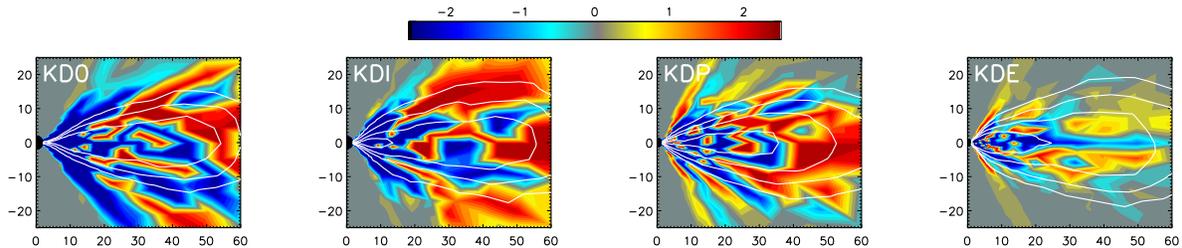}
    \caption{\label{KDmflux} 
     Color contours of the azimuthally-averaged mass flux $(\rho\,U^r)$ 
     time-averaged over the tenth orbit for the four
     high-resolution models. 
     In each plot the color scale is set to
     emphasize the outflows in the coronal envelope and the funnel-wall
     jet; the color scale multiplied by $10^{-7}$ for KD0, $10^{-6}$ for KDI
     and KDP, and $10^{-5}$ for KDE. 
     The main body of the disk is indicated by density contours
     of ${10}^{-2}$, ${10}^{-1.5}$, ${10}^{-1}$, and 
     ${10}^{-0.5}$ in units with ${\rho}_{max}(t=0)=1$. 
     The individual plots are labeled by model.  } 
\end{figure}

The coronal envelope features larger-scale, more regular radial motions 
than the main disk, which is dominated by turbulence.
These motions are most clearly visible in animations of poloidal slices of gas density, where gas is seen to blow back from the inner regions of the 
disk out along the disk surface. It is this ouflowing matter from the disk that fills the coronal envelope, and this outflow appears to be a
general outcome of turbulence-driven accretion (HB02). Figure \ref{KDmflux} 
shows the azimuthal-average of radial mass flux time-averaged over the tenth
orbit for each of the high-resolution models.  The range of the color
maps is chosen for each model to emphasize the radial motions in the coronal
envelope and the funnel-wall jet.  The color map saturates in the
main disk body, where turbulent fluctuations dominate.  The figure
shows that although the outflow remains irregular even in the coronal
envelope, its coherence length is larger than in the disk body.
In all four models, the radial flux in 
the coronal envelope is strongest near the surface of the disk,
and seems to originate at radii greater than $r \simeq 10$--$20\,M$,
with a tendency for the point of origin to move inward with
increasing $a/M$. 
Note that there is a key difference between the outward
coronal flows and the funnel-wall jets: the former are bound, while the latter are unbound ($-h\,U_t > 1$); in a global sense, the coronal flow
is therefore more circulatory than consistently outward.

Plots of the magnetic field structure in the coronal envelope
(not shown) also reveal a more regular structure to the magnetic
field than is the case in the disk body, but a less regular structure than
is found in the funnel-wall jet or the axial funnel.  

\subsection{The Inner Torus and Plunging Region}\label{innertorus}

An important objective for these numerical studies is to investigate the
nature of the inner edge of a black hole accretion disk.
The inner disk consists of two regions, which we have labeled the
inner torus and the plunging region.  The inner torus is located
where the accretion flow attains a local pressure maximum
just outside the marginally stable orbit.  In all four models, this
location is near $r \approx 1.6\,r_{ms}$.  Detailed time-histories of
average density (Fig. \ref{rhost}) show that the inner torus is a
time-varying structure that mediates the accretion and outflow in its
vicinity.  Figure \ref{rhoseq} shows that the density rises in
a relatively continuous (though time variable) way from the disk body
to the density maximum marking the inner torus, and then falls sharply 
toward smaller radii.  It is the region inside the inner torus and 
extending into the plunging region that contains the various
``inner edges'' defined by KH02. These are the locations where
turbulence gives way to streaming flow, the disk becomes optically thin,
significant local radiation ceases, and dynamical contact with the
outer flow can no longer be maintained.

The spacetime diagrams of density in Figure \ref{rhost} provide a good
deal of information on the dynamical behavior of the inner torus.
Clearly visible in the plots are small-scale, frequent accretion events
into the black hole from the inner edge of the inner torus.  When
accretion from larger radii isn't sufficient to immediately resupply
the inner torus after such an accretion event, the density in the inner
torus is greatly reduced.  This can be seen from $t \simeq 3$ to $t
\simeq 6$ and after $t \simeq 8.5$ in model KD0, from $t \simeq 5$ to
$t \simeq 7$ in model KDI, and from $t \simeq 7.5$ to $t \simeq 8.3$ in
model KDP. The inner torus in model KDE seems to grow progressively
with time, and is at its peak at the end of the simulation, which
suggests that the inner region in this model has not attained a
quasi-steady state even at this late time.

\begin{figure}[ht]
    \epsscale{1.0}
    \plotone{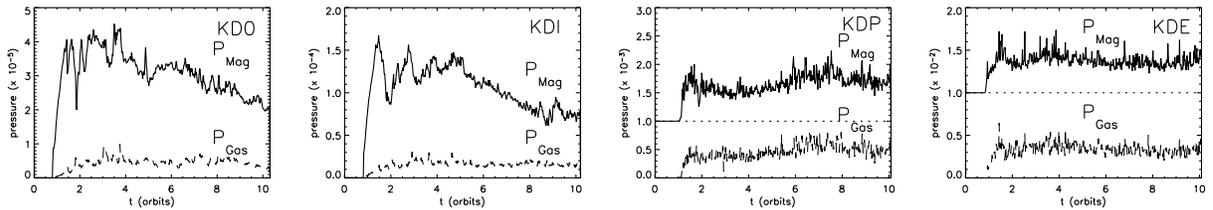}
    \caption{\label{KD_Pmg_rmin} 
     Gas and magnetic pressure history, $\langle{P_{gas}}\rangle$
     and $\langle{P_{mag}}\rangle$ at the inner radial boundary.
     For clarity, the $\langle{P_{mag}}\rangle$ curves for models KDP and 
     KDE are shifted vertically by $1.0 \times {10}^{-3}$ and 
     $1.0 \times {10}^{-2}$, respectively. The individual plots are 
     labeled by model.  Pressure is in code units, 
     as in Figure \ref{KD_Pmg_rpmax}. } 
\end{figure}

Although the main body of the disk remains gas-pressure dominated at late
time (Fig.~\ref{KD_Pmg_rpmax}), the magnetic pressure becomes
increasingly significant at small radius in all the models.  Figure
\ref{KD_Pmg_rmin} shows gas and magnetic pressure histories at the
inner radial boundary. Because gas pressure is largely confined to
the equatorial plane, the shell-integrated values of
gas pressure reflect conditions in this region. As for magnetic pressure,
it is distributed nearly spherically at small radii, so shell-integrated 
values of magnetic pressure show roughly a factor of two enhancement over 
typical values in the equatorial plane due to this more extended 
distribution. The magnetic pressure clearly dominates in
models KD0 and KDI, but magnetic and gas pressure are comparable in
models KDP and KDE.  In part, the difference is due to the relative
location of the inner radial grid boundary with respect to the
marginally stable orbit; the marginally stable 
orbit is comparatively closer to the inner boundary in the higher
black hole spin models. Between $r_{ms}$ and the horizon, the flow
accelerates inward, and the magnetic field evolves primarily by flux
freezing.  The net result is an increase in magnetic pressure relative
to the gas pressure.  The larger ratio of magnetic to 
gas pressure in models KD0 and KDI
may be at least partially a consequence of the greater separation
between $r_{ms}$ and the inner boundary.

\begin{figure}[ht]
    \epsscale{0.5}
    \plotone{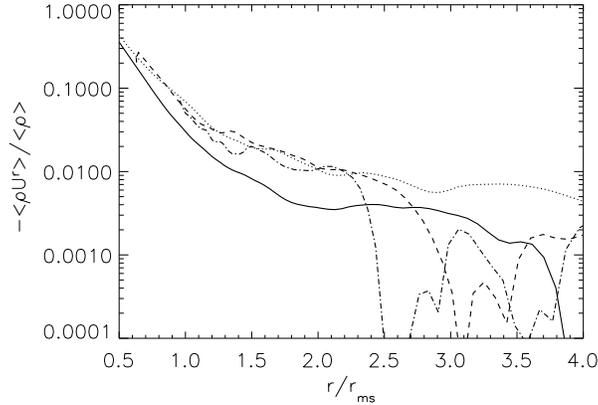}
    \caption{\label{KDvelplt} 
     Radial velocity profile as a function of $r/r_{ms}$, computed
     from $-\langle{\rho\,U^r}\rangle/\langle{\rho}\rangle$
     at the end time of the simulation. Four models shown, 
     KD0 (solid line), KDI (dotted line), 
     KDP (dashed line), and KDE (dash-dot line)} 
\end{figure}

As discussed in KH02, there are several possible ways to mark the inner
edge of a black hole accretion disk.  One of these is the turbulence
edge, the point where the flow makes the transition from
turbulence-dominated to plunging inflow past the marginally stable
orbit.  One indicator of this edge is the point where the radial infall
velocity begins to climb with decreasing $r$.  Figure \ref{KDvelplt}
shows the radial velocity profile, obtained from the history ratio
$\langle\rho\,U^r\rangle/\langle\rho\rangle$, and plotted as a function
of $r/r_{ms}$.  This plot clearly shows the transition to a smooth
accelerating inflow past the marginally stable orbit in all four
models.  The velocity begins to increase inside of about $2\,r_{ms}$,
much as it does in the pseudo-Newtonian calculation (cf. Fig.~3 of
KH02).

\begin{figure}[ht]
    \epsscale{0.5}
    \plotone{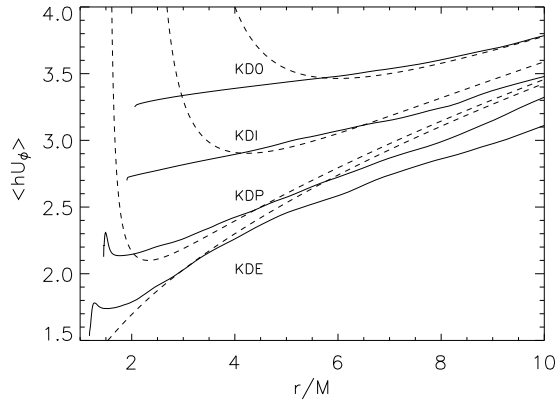}
    \caption{\label{KDuphi} 
     Density-weighted shell-average of 
     $\langle h\,U_\phi\rangle=\langle \rho\,h\,U_\phi\rangle/
      \langle \rho \rangle$
     time-averaged over the last orbit.  The dashed lines are
     the values of $U_\phi$ for a circular particle orbit.
     } 
\end{figure}

One motivation for the pseudo-Newtonian simulations of HK01 and HK02
was to investigate the degree to which magnetic stress continues within
the plunging region inside the marginally stable orbit.  Here we also
see evidence for this phenomenon.  Figure~\ref{KDuphi} shows the
density-weighted shell average of $hU_\phi$ in the models along with
the value of $U_\phi(r)$ for circular particle orbits in the equatorial
plane.  The KD0 and KDI models have the most coordinate distance
between $r_{ms}$ and the inner grid boundary and they show the effect
most clearly:  $hU_\phi$ continues to decline with an almost constant
slope.  The total additional decrease is modest, but does demonstrate
the continuing role of the magnetic stresses even inside the plunging
region.  In a subsequent paper we will study in greater detail the
magnitude of stresses connecting the plunging region to the disk
outside $r_{ms}$.

Inside the disk the instantaneous accretion rate, $\dot{M}=-\langle \rho
U^r\rangle$, at any one radius is highly variable.  This is a
consequence of the turbulence:  the velocity fluctuations are
significantly larger than the time-averaged radial drift velocity.
Averaging over time produces a more representative measure of $\dot M$,
but even then there are significant variations on longer timescales.
For example, animations of the disk density reveal local density maxima
that gradually inspiral through the main disk.  These maxima account
for the prominent accretion events seen in space-time diagrams.   In
addition to creating large variations in $\dot M$, these episodic
events feed bursts of gas into the coronal outflow and funnel wall
jets.  Computing the fractional rms accretion rate fluctuation as a
function of radius, we find that it is approximately $20\%$ in the
region of the inner torus and at smaller radii, but rises to values
several times greater at radii only twice as large.  The details of
these fluctuations will be studied in a later paper.

\begin{figure}[ht]
    \epsscale{0.5}
    \plotone{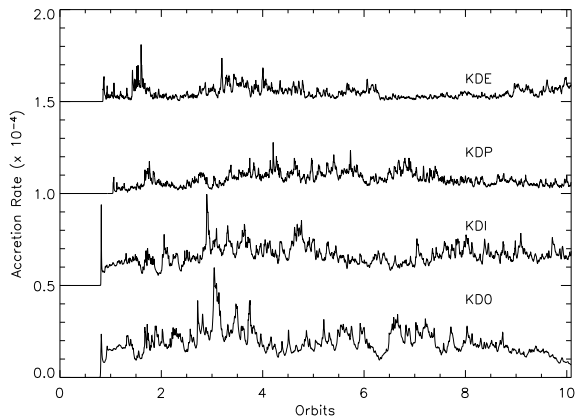}
    \caption{\label{KDmdot} 
     The accretion rate $\dot{M}=-\langle{\rho\,U^r}\rangle(r_{min})$ 
     through the inner radial boundary as a function of time 
     for the four high-resolution models.  
     The units are fraction of initial torus mass accreted per unit
     time $M$. For clarity, the KDI, KDP, and KDE curves are shifted 
     vertically by $0.5$, $1.0$, and $1.5$ $\times {10}^{-4}$,
     respectively.} 
\end{figure}

Some sense of the temporal fluctuations in $\dot M$ can be gleaned from
Figure \ref{KDmdot}, which shows the accretion rate through the inner
radial boundary as a function of time.  The units are the fraction of
initial torus mass accreted per $M$ of time.  Model KDP shows a slight
delay in the onset of accretion compared to the other simulations; this
is likely due to the higher $\beta$ that was used to initialize this
model.  Following the initial phase in which the MRI is established,
the rate of accretion, though variable, seems to settle to a mean value
which clearly decreases with increasing black hole spin, $a/M$.
Averaging in time over the second half of the simulations the mean
values are $1.78\times 10^{-5}$ for KD0, $1.61\times 10^{-5}$ for KDI,
$0.82\times 10^{-5}$ for KDP, and $0.43\times 10^{-5}$ for KDE.
Although the initial torus mass is larger for larger $a/M$, the
unnormalized accretion rates for KDP and KDE are still lower than that
of KD0.  The decrease of accretion rate with increasing $a/M$ was also
noted in DH03b, but the connection to black hole spin was not as
explicit.  In the SF models the ratio ${r_P}_{max}/r_{ms}$ varied
greatly over the range of models.  Here, with the initial pressure
maximum farther out, the variation of this ratio is less pronounced
across models, yet, again, the black hole spin is related to the
observed accretion rate.

\subsection{The Funnel-wall Jet}\label{jet}

Figure \ref{KDmflux} in \S\ref{corona} shows the azimuthally averaged
radial mass flux, $\overline{\rho\,U^r} (r,\theta)$, time-averaged over 
the tenth
orbit for each of the high-resolution models.  In each plot
isodensity contours are added to locate the main body of the disk and the 
inner torus.   These plots show the outflow along the funnel wall which we
have termed the ``funnel-wall jet.''  This jet is particularly
prominent in the plots for models KDP and KDE.  The funnel-wall jet
originates at the boundary between the coronal envelope and the funnel,
at the location of the inner torus.  In model KD0, where the inner
torus is the least prominent, the outflow is extremely weak.  As the
black hole spin increases, the inner torus shifts inward, becoming
increasingly prominent.  The base of the jet moves inward with the
inner torus; for model KDI, the base of the jet lies just outside the
static limit, while in models KDP and KDE, it is inside the
ergosphere.  In addition, the strength of the jet compared to the
radial mass flux in the main disk body also increases with
increasing $a/M$.

The property distinguishing the jet is that the
outflowing material within it is unbound, i.e., $-h\,U_t > 1$.  For
purposes of a preliminary examination of jet properties, we have
studied the jet in model KDP in greater detail and found that the sum
of the forces acting on it at large $r$ are decelerative (gravity).
This suggests that the jet is driven impulsively at or near the launch
points, and that its subsequent motion is ballistic.

\begin{figure}[ht]
    \epsscale{0.7}
    \plotone{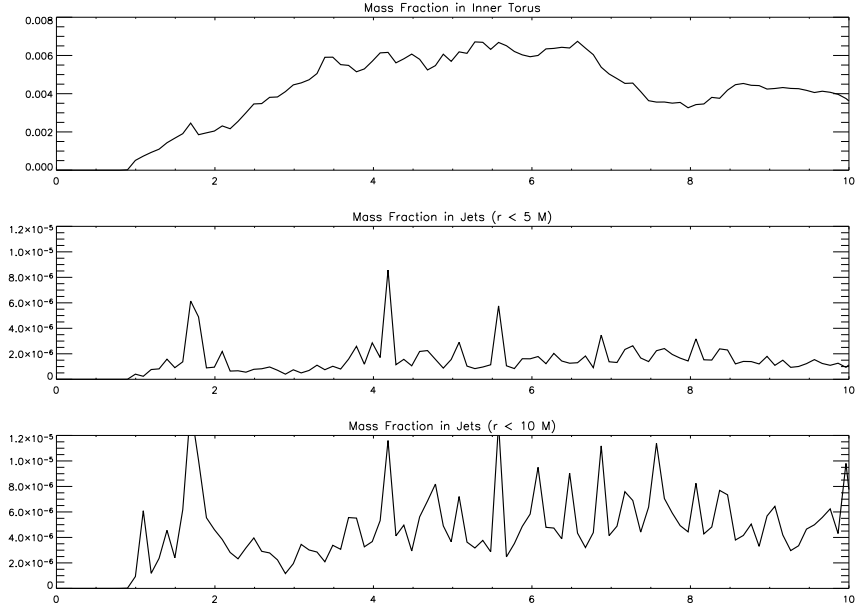}
    \caption{\label{jetmass} 
     Rest mass (as a fraction of total disk mass)
     in the inner torus, defined as mass inside $r = 6\,M$ (top plot), 
     and in the funnel-wall jet
     inside $r < 5\,M$ (middle plot), and inside $r < 10\,M$ (bottom
     plot) as a function of time for model KDP.
     The data for this plot were extracted
     from the periodic dumps of the code variables taken 
     every $80\,M$ 
     in time.  The jet is defined as material that is both unbound 
     and outbound (see text).
} 
\end{figure}

Animations of the near-hole region reveal that the jet is not a uniform
outflow, but is driven in episodic bursts that appear at a frequency
comparable to that of the accretion streams that feed the inner torus.
To illustrate this, Figure \ref{jetmass} shows a series of restricted
rest mass histories ($\left[\rho\,U^t\right]$) for model KDP. The top
curve shows the total rest mass within the inner torus as a function of
time (as a fraction of the initial disk rest mass).  The rest mass of
the inner torus is computed by restricting $\left[\rho\,U^t\right]$ to
$r < 6\,M$.  The two curves below this show the fraction of rest mass
in the jets for $r < 5\,M$ and $r < 10\,M$.  The mass in the jet is
extracted from the numerical data using two criteria:  gas deemed to be
in the jet is unbound, $-h\,U_t > 1$, and outbound, $\rho\,U^r >
({\rho\,U^r})_{min} > 0$, where $({\rho\,U^r})_{min}$ is a mass flux
threshold chosen to exclude the more tenuous funnel outflow.  The cut 
was set at 0.5 \% of the maximum instantaneous unbound mass flux.

Figure \ref{jetmass} shows a steady growth of the mass of the inner
torus from $t \simeq 1$ to $t \simeq 3.5$ orbits, followed by a decline
from $t \simeq 7.5$ to $8.3$ orbits.  This behavior can also be seen in
the spacetime diagram, Fig.~\ref{rhost}.  Figure \ref{jetmass} shows
that significant mass is injected into the jet inside $r < 5\,M$ at
only three instants, $t \simeq 1.8$, $4.2$, and $5.6$ orbits.  These
features occur just after local, albeit small, upsurges in inner torus
mass.  They also seem to coincide with accretion events from the inner
torus into the black hole (Fig. \ref{rhost}).  At other times, the $r <
5\,M$ curve suggests that matter is fed episodically at a low level
near the origin of the jet. It appears, however, that a greater amount
of matter is injected into the jet between $5\,M < r < 10\,M$ (bottom
curve).  The large spike in mass at $t \simeq 1.2$ orbits seen in the
bottom curve is a feature related to the outflows that occur during the
initial inflow of material from the torus; the funnel wall jet is not
yet established at this time.  Together, the two jet mass curves
suggest that, although the jet's origin lies near the black hole, a
greater portion of
the mass in the jet is injected episodically between $r \approx
5\,M$ and $10\,M$.  A more extensive analysis of the funnel-wall jet
will be presented in a subsequent paper.

\subsection{The Axial Funnel}\label{funnel}

The funnel is the centrifugally-evacuated region near the north and
south polar axes. The funnel is magnetically dominated, and contains a
very tenuous gas which is predominantly moving outward at relativistic
velocities ($V^r \simeq 0.95\,c$).  This thin wind is thermally driven;
the tenuous gas is very hot with enthalpy $h \gg 1$.  The high
temperatures result from shocks driven into the funnel by the accretion
disk and the corona.  Typical gas densities are above the numerical
floor, but several orders of magnitude below the density in the main
disk body, and one to two orders of magnitude below that of the
funnel-wall jet.  The small amount of gas present in the funnel has
negligible angular momentum, and the edge of the funnel coincides with
a very sharp gradient in specific angular momentum that divides coronal
gas from funnel gas.  Even though there is vigorous outward transport of
angular momentum, the gas in the disk retains more
than enough specific angular momentum to be excluded from the funnel.
In the KDP and KDE simulations some higher angular momentum gas does
reach the region near the axis, but only in the region very close to
the black hole.

Magnetic field, on the other hand, is not so excluded, and the funnel
is filled with a predominantly radial field.  Because the gas
density levels are low, the Alfv\'en speeds here are high.  The
magnetic pressure, however, is not extraordinarily large; even at the
base of the funnel ($r\simeq 5\,M$) it is comparable to $P_{gas}$ at the
original torus pressure maximum.  As shown in Figure \ref{KD0pazavg}, the
magnetic pressure is nearly spherically symmetric within the funnel;
gas pressure here is negligible.  Gas pressure dominates within the
turbulent disk, and drops abruptly across the funnel wall, but
the total pressure is smooth and continuous.

\subsection{Global Quantities}\label{globalqty}

Aspects of the overall evolution of the disk can be characterized by
the time evolution of the total mass ($M$), energy ($E$), and momentum
($L$), computed using equations (\ref{Mflux})---(\ref{Jflux}).  Table
\ref{Mtotals} lists the initial ($Q_{\rm{0}}$) and final ($Q_{\rm{f}}$)
values on the grid, as well as the cumulative outflow of each quantity
through the inner ($\Delta Q_{\rm{i}}$) and outer ($\Delta Q_{\rm{o}}$)
boundaries, for the four disk models at both low- and high-resolution.
These values were computed every $1\,M$ in time, corresponding to
roughly once every 100 time steps.  The boundary fluxes were obtained
approximately from state variables at the end of a timestep, rather
than from the flux values computed by the numerical algorithm
within a time step.  Since this procedure is approximate, the sum of
the outflows and final values on the grid need not match perfectly:
for example, the rest mass totals are low compared to the initial
values by about 2\%.

\begin{table}[ht]
\caption{\label{Mtotals}
  Conserved Quantities in Low- and High-Resolution Models}
\begin{tabular}{lrrrrrrrrrrrrrr}
 & & & & & & & & & & & & &\\
\hline
Model & $M_{\rm{0}}$ & $M_{\rm{f}}$ & $\Delta M_{\rm{i}}$ &
 $\Delta M_{\rm{o}}$  & $ {\Delta M_{\rm{i}} \over M_{\rm{0}}}$ 
 & $E_0$ & $E_{\rm{f}}$ & $\Delta E_{\rm{i}}$ & 
 $\Delta E_{\rm{o}}$ 
 & $L_0$ & $L_{\rm{f}}$ & $\Delta L_{\rm{i}}$ & 
 $\Delta L_{\rm{o}}$  &
 $ {\Delta L_{\rm{i}} \over \Delta M_{\rm{i}}}$\\
\hline
\hline
KD0lr      & 156   & 140   &  13.9  & 0.93 & 0.089 
           & 154   & 138   &  13.9  & 0.99  
           & 902   & 847   &  44.0  & 4.42 & 3.17 \\
KD0        & 156   & 130   &  21.6  & 1.10 & 0.138 
           & 154   & 129   &  20.7  & 1.17  
           & 902   & 817   &  68.8  & 5.66 & 3.19 \\
KDIlr      & 258   & 232   &  19.7  & 3.45 & 0.076 
           & 255   & 229   &  20.1  & 4.07 
           & 1488  & 1407  &  51.8  & 20.6 & 2.63 \\
KDI        & 258   & 219   &  31.5  & 2.82 & 0.122 
           & 255   & 217   &  30.5  & 3.65 
           & 1489  & 1378  &  82.2  & 16.7 & 2.61 \\
KDPlr      & 291   & 277   &  10.7  & 1.57 & 0.037 
           & 286   & 273   &  11.8  & 2.78 
           & 1652  & 1622  &  19.3  & 6.77 & 1.80 \\
KDP        & 291   & 269   &  16.5  & 2.48 & 0.056 
           & 286   & 266   &  16.2  & 3.76 
           & 1652  & 1606  &  21.1  & 11.6 & 1.28 \\
KDElr      & 392   & 370   &  8.05  & 11.8 & 0.020
           & 386   & 364   &  15.6  & 21.5
           & 2254  & 2189  &  -0.88  & 61.1 & -0.11 \\
KDE        & 392   & 364   &  13.8  & 11.3 & 0.035 
           & 386   & 359   &  17.9  & 21.5 
           & 2255  & 2178  &  5.79  & 66.6 & 0.42 \\
\hline
\end{tabular}
\end{table}

Table \ref{Mtotals} shows that accretion into the black hole dominates
over flow through the outer boundary, except for the KDE models.
Generally, it is desirable to avoid significant losses through the
outer boundary since such losses could remove dynamically important
material from the causal domain of the simulation.  Such losses are
minimized by locating the outer boundary at large enough
radius that the outer part of the disk does not reach it over the
course of a typical evolution.  The exception to this is model KDE,
where the higher losses through the outer boundary result
from the larger relative size of the initial torus.

The largest total rest mass accreted as a fraction of the initial mass is
14\% for model KD0.  The accreted mass decreases with increasing $a/M$,
and the low-resolution simulations consistently yield a lower accreted
mass than their high-resolution counterparts.  In the pseudo-Newtonian
simulation of HK02, 16.6\% of the initial torus mass is accreted in 10
orbits of time at its initial pressure maximum ($R=20M$).  Compared to
the SF0 and SFP models of DH03b, the fraction of disk mass accreted
into the black hole in the KD models is smaller, even though the KD
models are run for a longer absolute time.  The SF models feature a
strong transient accretion flow during the early evolution of the
initially constant angular momentum torus.  This contributes to the
higher accreted mass in these models.

Over the course of the evolution the average specific energy $E/M$
increases with time as the gas on the grid becomes less bound.
The greatest change in average specific energy occurs for the low $a/M$
models, $0.23$ \% for KD0 and $0.25$ \% for KDI. Smaller changes
occur for the high $a/M$ models, $0.11$ \% for KDP and $0.01$ \% for
KDE. In the last case the loss of less-bound material through the outer 
boundary seems to nearly compensate for the inflow 
of more-bound material into the black hole.

The average specific angular momentum on the grid, $L/M$, also
increases with time.  For all models except KDE, more low angular
momentum mass has entered the hole than high angular momentum material
has exited through the outer boundary.  There is considerable variation
between values obtained in the high- and low-resolution models of KDP
and KDE.  In particular, model KDElr shows a negative $\Delta
L_{\rm{i}}$, which is likely an artifact due to an under-resolved
plunging region.   However, the high-resolution models do show an
interesting overall trend:  the specific angular momentum accreted into
the hole, ${\Delta L_{\rm{i}}/\Delta M_{\rm{i}}}$, decreases with
increasing $a/M$ compared to the marginally stable value, $l_{ms}$.  On
the other hand, the values of $\langle h\,U_\phi \rangle$
(Fig.~\ref{KDuphi}) are below the marginally stable value of $U_\phi$
for models KD0 and KDI where there is an extended plunging region, but
comparable to ${(U_\phi)}_{ms}$ for models KDP and KDE, where $\langle 
h\,U_\phi \rangle \approx 2.1$ and $1.5$, respectively.  The stresses and 
fluxes within the plunging regions will be examined in greater detail in a 
subsequent paper.

\section{Discussion}

We have computed a set of three-dimensional MHD accretion simulations
for a range of Kerr black hole spin parameters.  These models are
general relativistic versions of previous pseudo-Newtonian simulations
(HK01, HK02) that studied the evolution of a torus with a
near-Keplerian angular momentum distribution and weak poloidal loops of
magnetic field.  The accretion flows are non-radiative, and evolve from
a fixed amount of initial mass, which increases with increasing $a/M$.
In this paper we present an initial survey of the results,  describing
the overall structure of the accretion flow that develops.  We label
and describe a set of five specific features in the accretion flow.
These are the turbulent, dense, main disk body with a roughly constant
$H/R$, the inner disk consisting of an inner torus and a plunging
inflow, a magnetized coronal envelope surrounding the disk, a jet-like
outflow along the funnel wall, and an evacuated axial funnel.

In our preliminary analysis we are looking for ways in which the spin
of the black hole affects the overall evolution.  One such effect that
immediately emerges is a decrease in the accretion rate into the hole
with increasing spin parameter $a/M$.  This was also seen in the
previous study of constant-$l$ tori in the Kerr metric (DH03b), where   
the role of black hole spin was
partially explored, but the issue of accretion rate and its relation to
$a/M$ was somewhat obscured by the choice of initial conditions.  The
initial tori had inner boundaries that were closer to the black hole
than for the tori used here, and the range of spin parameters yielded a
substantial variation in the ratio ${r_P}_{max}/r_{ms}$.  In the KD
models, the initial tori are located at larger radii compared to
$r_{ms}$, and the variation in the ratio of initial pressure maximum to
marginally stable orbit is less pronounced.  There are several ways
that increased black hole spin can influence the accretion rate.  One
is that the inner torus, located near $r_{ms}$, is thicker and hotter
with increasing black hole spin.  In the absence of cooling, this high
pressure inner torus could choke off the accretion flow.  Another way
that the accretion rate could be influenced is through the
$a/M$-dependent shape of the effective potential near $r_{ms}$. A
third way,
suggested by the large electromagnetic angular momentum flux seen in KDP
and KDE, is for magnetic torques driven by the black hole's rotation to
supply some of the outward angular momentum flux through the disk body.
These dynamical effects will be investigated in greater detail in a
subsequent paper.

In all models the accretion rate is highly time- and space-variable,
and this, in turn, makes the characteristics of the inner torus vary in
time as larger or smaller amounts of gas are brought into the near-hole
region from the extended disk.  The inner torus therefore exerts an
important influence on the accretion flow into the black hole, acting
as a temporary reservoir that feeds some matter through the potential
``cusp'' and into the plunging region, while also sending mass outward
along the funnel wall.  This latter feature, the funnel-wall jet, is
launched somewhat sporadically from a point in the coronal envelope
above the inner torus.  The jet flows outward along the boundary
between the coronal envelope and the axial funnel.  The intensity of
the jet increases with black hole spin, as the origin of the jet moves
gradually deeper into the potential well of the black hole.

As in the previous pseudo-Newtonian simulations, we find that
significant stress continues to operate well inside the
marginally stable orbit.  As a result, the angular momentum accreted
per unit mass is less than the specific angular momentum of matter
following the marginally stable orbit, in contrast to the expectations
of the classical Novikov-Thorne model (Novikov \& Thorne 1973).  When
the black hole is rotating relatively slowly or not at all (KDI, KD0),
the difference between the accreted $l_{in}$ and $l_{ms}$ is noticeable
but modest, $l_{in}/l_{ms}\sim 85\%$.  On the other hand, when $a/M =
0.9$ (KDP), the angular momentum per unit mass carried into the hole is
about $50\%\ l_{ms}$, and when $a/M=0.998$ (KDE), the value is only
$20\%$.  While these specific numbers should be regarded with caution, as
there is considerable variation with different grid resolution, the
trend with increasing $a/M$ seems more robust. 
In the lower-spin simulations (KD0 and KDI), the net angular momentum
accreted by the black hole is dominated by the part directly associated
with matter (i.e., near the inner boundary ${T^r}_\phi \simeq h\rho U^r
U_\phi$) ; however, as the spin increases, the angular momentum
associated with the magnetic field (i.e., the terms 
$\|b\|^2 U^r U_\phi-b^r b_\phi$ in the stress tensor) become increasingly
important.  In fact, most of the reduction in accreted specific angular
momentum seen in the higher-spin simulations is due to
{\it negative} magnetic angular momentum flux.  It is interesting to
note that in the KDE simulation, this effect is so strong that there 
could be spin-down of the black hole (if the background metric were
not fixed), as the angular momentum accreted
per unit mass is less than the angular momentum per unit
mass stored in the black hole.

The models discussed here represent one more step in a program to
pursue increasingly realistic simulations of magnetically driven
accretion flows in black hole spacetimes. With the availability of
significant computing power, spatial resolution is less of a limiting
constraint than it was just a few years earlier; however,
higher-resolution simulations than were carried out here remain 
necessary to address the issue of numerical convergence.  
For the present study we have run each model at two resolutions.  We
see systematic, resolution-dependent differences in such quantities as
accretion rate through the inner radial boundary, and the specific
angular momentum and energy of that accretion flow.  These models also
were run on a restricted azimuthal grid of $\pi/2$.  As noted in DH03b,
there are differences with a full $2\pi$ model, including a reduction
in the variability of the accretion rate, and an increase in the
average accretion rate.
Resolution through the main disk body is also a critical issue,
since the modes that sustain the MRI must be adequately resolved for
turbulence to sustain accretion throughout the simulation. Experiments
have shown that the lower-resolution (lr) grids used here are near the
lower limit in this regard.

While these considerations certainly argue for the desirability of
increased resolution, the grids that we have been able to use here have
allowed us to probe for the first time some of the time-dependent
dynamics of accretion flows down to near the event horizon of rotating
black holes, within the limits of the Boyer-Lindquist coordinate
system.  In subsequent papers in this series we will to look more
deeply into these models, including
the role of stress inside the marginally stable orbit and
the apparent influence of black hole spin, as well as investigate more
closely the details in the regions of the flow identified here.

\acknowledgements{This work was supported by NSF grant
AST-0070979 and NASA grant NAG5-9266 (JPD and JFH),
and by NSF Grant AST-0205806 (JHK).  We thank Charles Gammie 
and Shigenobu Hirose for
valuable discussions related to this work.
The simulations were carried out on the
Bluehorizon system of NPACI.}


\end{document}